\documentclass[fleqn,12pt,twoside]{article}
\usepackage[headings]{espcrc1}

\readRCS
$Id: espcrc1.tex,v 1.2 2004/02/24 11:22:11 spepping Exp $
\usepackage{graphicx}
\usepackage[figuresright]{rotating}

\newcommand{\AmS}{{\protect\the\textfont2
  A\kern-.1667em\lower.5ex\hbox{M}\kern-.125emS}}

  \def\dis{\displaystyle}   \def\beq{\begin{eqnarray}}
  \def\eeq{\end{eqnarray}}    
  \def\be{\begin{equation}}  \def\ee{\end{equation}}
\title{Multiquark states as di-hadronic molecules}

\author{Ajay Kumar Rai\address[MCSD]{Department of Physics, Sardar Patel University,
Vallabh Vidyanagar, \\ Gujarat-388 120, INDIA.}$^{,b}$%
                J N Pandya\address{Applied Physics Dept., Faculty of Technology and Engineering,\\
 M S University of Baroda, Vadodara, Gujarat-390 001, INDIA.}
        and
         P C Vinodkummar  \addressmark[MCSD]}

\runtitle{Multiquark states as di-hadronic molecules.}
\runauthor{Ajay Kumar Rai et al.}
\begin{document}
\maketitle
\begin{abstract}
Multiquark systems such as tetraquarks, pentaquarks and hexaquarks
states are studied as di-hadronic molecules in a nonrelativistic
model. The masses of several di-hadronic states using a molecular
interaction provided by asymptotic expression of the confined gluon
exchange potential are computed. The exotic states such as f$_0$
(0.982), f$_2$ (1.565), f$_2$ (1.950), X(3.87), D$_{sJ}$ (2.317,
2.460, 2.632), $\psi$(4.040) etc are identified as the low lying
di-mesonic states, and $\Theta^+$ (1.54) as a $K-N$ molecular state.
\end{abstract}
\section{Introduction}
The exotic states include those which do not fit in the $q \bar q$
and $qqq$ states. So they include the tetra quark states, penta
quark states, hexa quark states, hybrids such as $q \bar q g$ and
$gg$ or $ggg$ glueball states. The multi-quark system has previously
been studied in the framework of Bag model \cite{RLJaffe1977} and
nonrelativistic potential models \cite{JWeinstein}. But there exist
little succuss towards understanding the tetra-quark and penta-quark
states as di-hadron molecules due to nonperturbative nature of QCD
at the hadronic scale. It is widely believed that the QCD sum rule
and the instanton based models are capable nonperturbative methods
to extract the properties of hadrons \cite{Diakonov,EOset1997} but
there exist little works on multiquark states, especially their
systematic calculations \cite{Junsuiman}. Recently, several new
states X(3.872), D$_{sJ}$ (2.317, 2.460, 2.632),Z(3.931), Y(3.943),
Y(4.260) etc were observed in the experimental
investigations of charmed mesons. 
 These experimental observations have generated renewed interest in
their theoretical interpretation
\cite{Debert2006,Marina2006,Rosina2004}. Following the molecular
interpretation of these states \cite{Colaggelo2004} we study these
systems of multiquark states as di-hadronic molecules. For the
binding energy of the di-hadronic state, we consider a large r ($r
\rightarrow \infty$) limit of the confined gluon propagator that
appear in the COGEP \cite{Ajay2006,JNPandyap2001}. This residual
confined gluon interaction conceptually is similar to the Van-der
Waal interaction of the atomic molecules.
\section{Masses of Di-hadrons as molecular states}
The low-lying di-hadronic molecular system consisting of either
di-meson tetra quark states, meson-baryon pentaquark states and
baryon-baryon hexa quark states are studied non-relativistically.
Here the di-hadronic Hamiltonian is assumed as, \be
\label{eq:hammol} H = M + \frac{P^2}{2 \mu} + V(R_{12})+ V_{SD}(S_1
S_2) \ee where $M=m_{h_1}+m_{h_2}$, $m_{h_1}$ and $m_{h_2}$ are
masses  of the hadrons, $\mu$ is the reduced mass, $P$ is the
relative momentum of the two hadrons and $V(R_{12})$ is the residual
(molecular) interaction potential between the two hadrons. We
express it as the asymptotic expression ($r\rightarrow \infty$) of
the confined one gluon exchange interaction (COGEP) given by
\cite{Ajay2006,JNPandyap2001}
  \be V(R_{12})=\frac{-k_{mol}}{R_{12}} e^{-C^2 \
R_{12}^2/2}\label{eq:confgluon1}\ee where $k_{mol}$ is the residual
strength of the strong interaction coupling and $C$ is the effective
colour screening parameter of the confined gluons. Using a trial
wave function given by \be \label{eq:wavfu} \psi(R_{12})= \left(4
\frac{\Omega^{3/2}}{\sqrt{\pi}} \right)^{1/2} e^{-\Omega \ \
R^2_{12}/2}\ee the ground state energy is obtained by minimizing the
expectation value of H as \be E(\Omega)=\langle
\psi(R_{12})|H|\psi(R_{12})\rangle {\sf \ \ \ and \ \ \  }
\frac{dE(\Omega)}{d \Omega}=0 \label{eq:eomega}\ee With the
resultant variational parameter $\Omega$ satisfying
Eqn(\ref{eq:eomega}) the mass of the low lying di-hadronic state is
obtained as \be E(\Omega)= M + \dis\frac{3\Omega}{4 \mu} -
\dis\frac{4 k_{mol}\Omega^{3/2}}{c^2 + 2 \Omega} + \frac{8}{9}
\frac{ \alpha_s}{m_{h_1} m_{h_2}}
 \ \vec S_1 \ \cdot  \vec S_2 \  |\psi(0)|^2\label{eq:varmass1}\ee
Where, we have added the spin-hyperfine contribution separately. The
binding energy of the di-mesons is obtained by subtracting the rest
mass energies of the respective constituent hadrons as
$BE=|m_{h_1}+m_{h_2}-E|$.\\
\indent We have computed the masses and binding energies of the
di-hadronic systems by choosing the residual interaction strength
$k_{mol}$ as the strong interaction strength, $\alpha_s$ at the
respective hadronic scale. Accordingly for light flavour hadrons (u,
d and s sector), $k_{mol}$=0.7 and for the charmed sector
$k_{mol}$=0.45 are considered. The gluon confinement parameter C
=0.1 GeV for combinations of u, d and s quarks and C=0.250 GeV for c
quark combinations and the experimental masses of the constituting
hadrons  \cite{ParDgroup} are employed in the present computations.
\begin{table}
\caption{Low-lying masses of Multiquarks as di-hadronic molecule}
\label{ta:extet}
\begin{center}
\begin{tabular}{lccccccc}
\hline
Systems&J$^{PC}$&$\Omega$&$\psi$&BE&Mass&Expt\cite{ParDgroup}&Others[Ref]\\
$h_1-h_2$&&$GeV^2$&$GeV^{3/2}$&$GeV$&$GeV$&$GeV$&$GeV$\\
\hline
$\pi$-$\pi$&$0^{++}$&0.0032&0.0203&0.017&0.297&& \\
$\pi$-$K$&$0^{++}$&0.0131&0.0582&0.025&0.659&$f_0(0.600)$& \\
$\pi$-$\rho$ &$1^{+-}$&0.0155&0.0659&0.024&0.935&&\\
$K$-$K$&$0^{++}$&0.0603&0.1828&0.004&0.992&$f_0(0.980)$&0.987\cite{EOset1997}\\
$\pi$-$K^*$ &$1^{+-}$&0.0161&0.0680&0.023&1.055&$h_1(1.170)$&\\
$\rho$ -$K$&$1^{+-}$&0.0868&1.3512&0.004&1.261&$b_1(1.235)$&\\
$K$-$K^*$&$1^{+-}$&0.0961&0.2593&0.006&1.380&$h_1(1.380)$&\\
\hline
&$0^{++}$&0.1380&0.3401&0.196&1.346&$f_0(1.370)$&\\
$\rho$ -$\rho$&$1^{++}$&&&0.105&1.437&$f_1(1.420)$&\\
&$2^{++}$&&&0.076&1.618&$f_2(1.565)$&\\
\hline
&$0^{++}$&0.1577&0.3759&0.210&1.453&$a_0(1.450)$&\\
$\rho$-$K^*$&$1^{++}$&&&0.114&1.549&&\\
&$2^{++}$&&&0.078&1.741&$f_0(1.710)$&\\
\hline
&$0^{++}$&0.1822&0.4189&0.228&1.556&&\\
$K^*$-$K^*$ &$1^{++}$&&&0.125&1.659&&\\
&$2^{++}$&&&0.081&1.865&$f_2(1.950)$&\\
\hline
$\pi$-$D$&0$^{++}$&0.0186&0.0757&0.022&2.027&&\\
$\pi$-$D^*$&$1^{+-}$&0.0188&0.0762&0.022&2.169&&\\
$K$-$D$&0$^{++}$&0.1415&0.3465&0.015&2.344&$D^+_{sj}(2.317)$&\\
$K$-$D^*$&$1^{+-}$&0.1455&0.3539&0.016&2.485&$D^+_{sj}(2.460)$&\\
$\rho$ -$D$&$1^{+-}$&0.2684&0.5602&0.033&2.603&&\\
$K^*$-$D$&$1^{+-}$&0.3265&0.6489&0.039&2.718&&\\
\hline
&$0^{++}$&0.2795&0.5775&0.235&2.543&&\\
$\rho$ -$D^*$&$1^{++}$&&&0.134&2.644&&\\
&$2^{++}$&&&0.064&2.845&&\\
\hline
&$0^{++}$&0.3420&0.6718&0.158&2.624&$D_{sJ}(2.632)$&\\
$K^*$-$D^*$&$1^{++}$&&&0.040&2.741&&\\
&$2^{++}$&&&0.077&2.976&&\\
\hline
$D$-$D$&0$^{++}$&0.3568&0.6935&0.008&3.738&&3.723 \cite{Lmaiani2006}\\
$D$-$D^*$&$1^{+-}$&0.3810&0.7285&0.006&3.878&X(3.870)&3.876 \cite{Rosina2004}\\
\hline
&$0^{++}$&0.4081&0.7670&0.084&3.930&&\\
$D^*$-$D^*$&$1^{++}$&&&0.040&3.974&&\\
&$2^{++}$&&&0.048&4.062&$\psi(4.040)$&3.968 \cite{Debert2006}\\
\hline
K-N&$(1/2)^-$&0.1910&0.4340&0.106&1.540&$\Theta^+$(1.540)&1.54$\pm$ 0.01 \cite{TNakano} \\
$K$-$\Sigma$&$(1/2)^-$&0.2212&0.4845&0.102&1.785&&\\
K-$\Delta$&$(3/2)^-$&0.2255&0.4916&0.011&1.737&&\\
$K^*$-N&$(1/2)^-,(3/2)^-$&0.3663&0.7073&0.073&1.905&&\\
\hline
N-N &$0^+,1^+$&0.3843&0.7332&0.039&1.918&&\\
N-$\Delta$&$1^+,2^+$&0.4836&0.8712&0.053&2.225&&2.17 \cite{HGarcilo1999}\\
$\Delta$-$\Delta$&$0^+,1^+,$&0.6290&1.0610&0.032&2.496&&2.46 \cite{HGarcilo1999}\\
&$2^+,3^+$&&&&&\\
\hline
\end{tabular}
\end{center}
\end{table}

\section{Conclusion and Discussion:}
We have computed the low-lying masses of the multiquark states by
considering them as di-hadronic molecules. The inter-hadronic
interaction has been taken as that due to the asymptotic expression
of the confined one gluon exchange interaction among the quarks
\cite{JNPandyap2001}. While the hadronic masses have been taken from
the experimental results, the Van-der-Waal like of di-hadronic
binding energies have been computed using the variational approach
with the inter-hadronic interaction strength, $k_{mol}=\alpha_s$,
the strong interaction running coupling constant at the respective
hadronic scale.\\
\indent Our predictions shown in Table-\ref{ta:extet} are compared
and identified with some of the experimentally known exotic
hadronic states. These exotic states are those whose spin parity
do not match with the expected quark anti-quark structure for
mesons and 3-quark structure of baryons. Accordingly, the
pseudoscalar di-mesonic and vector-vector di-mesonic combinations
will have the parity and charge conjugation PC as ++ while for the
combinations of pseudoscalar-vector di-meson state will have PC
value + -, as shown in Table-2. The experimental candidates with
the predicted $J^{PC}$values are chosen for comparison.\\

\indent We found many $0^{++}$ di-mesonic states in the energy range
0.297 GeV to 1.865 GeV at the light flavour (u, d, s) sector. Many
of these states may be identified with experimentally known exotic
mesonic states. Accordingly we identify h$_1$(1.170 GeV) 1$^{+-}$
and f$_0$(0.980 GeV) 0$^{+-}$ states as the $\pi-K^* $ and K-K
di-mesonic states. Other di-mesonic states such as
$D^+_{sj}(2.317)$, $D^+_{sj}(2.460)$, D$_{sJ}$(2.632 GeV), X(3.870
GeV) and $\psi$(4.040 GeV) are identified here as $K-D$,
$K-D^*$,$K^*-D^*$, $D-D^*$ and $2D^*$ di-mesonic states
respectively. Among many combinations of mesons-baryon and di-baryon
molecular states which can be studied, only few low lying state are
presented here in Table-1. Though the interpretation of
$\Theta^+(1.54)$ is still doubted, we identify it as K-N molecular
state. Many of the other predicted di-hadronic states could be
experimentally identified.

\end{document}